\documentclass[super]{rsc-layout}
\usepackage[utf8]{inputenc} % support UTF-8 in bib file

\usepackage{amsmath,amssymb,graphicx,textcomp}
\usepackage{cleveref} % clever referencing
\usepackage{verbatim}

\usepackage{times}
\newlength\figwidth
\usepackage[usenames,dvipsnames]{xcolor}

\usepackage{bm}

\newcommand{\av}[1]{{\ensuremath{\left\langle\vphantom{1^2} #1 \right\rangle}}}
\newcommand\diff{\mathrm{d}}
\renewcommand{\vec}{\bm}
\renewcommand\leq{\leqslant}
\renewcommand\geq{\geqslant}

\DeclareMathAlphabet{\mathlm}{OML}{lmm}{m}{it}
\newcommand\vel{\ensuremath{\mathlm{v}}}

\newcommand\eps{\ensuremath{\varepsilon}}
\newcommand{\msd}{\ensuremath{\delta r^2}}

\newcommand{\tracer}{\ensuremath{_\text{WCA}}}
\newcommand{\matrx}{\ensuremath{_\text{core}^\text{WCA} }}
\newcommand{\WCA}{\ensuremath{_{\text{WCA}}}}

\newcommand{\core}{\ensuremath{_{\text{\!core}}}}
\newcommand{\crit}{\ensuremath{_{c}}}
\newcommand{\reduced}{\ensuremath{^*}}
\newcommand{\cut}{\ensuremath{_\mathrm{cut}}}

\newcommand{\BH}{\ensuremath{^\text{BH}}}
\newcommand{\perc}{\ensuremath{_\text{perc}}}
\renewcommand{\th}{\ensuremath{_\text{th}}}
\newcommand{\eff}{\ensuremath{_\text{eff}}}

\makeatletter
\newcommand\dash{\penalty\@M-\hskip\z@skip}
\makeatother

\usepackage[normalem]{ulem}

\newcommand\hide[1]{\textcolor{red}{}}

\usepackage{layouts}
\usepackage{multicol}
\usepackage{color}
\usepackage{listings}

\begin{document}

\title{Rounding of the localization transition in model porous media}

\author{%
Simon K. Schnyder$^\text{a}$, Markus Spanner$^\text{b}$, Felix H\"ofling$^\text{c}$, Thomas Franosch$^\text{d}$, and J\"urgen Horbach$^{\text{a},\dagger}$
}

\abstract{
The generic mechanisms of anomalous transport
in porous media are investigated by computer simulations of two-dimensional model systems. 
In order to bridge the gap between the strongly idealized Lorentz model and realistic models of porous media, two models of increasing complexity are considered: a cherry-pit model with hard-core correlations as well as a soft-potential model.
An ideal gas of tracer particles inserted into these
structures is found to exhibit anomalous transport which extends up to several decades in time. Also, the self-diffusion of the tracers becomes suppressed upon increasing the density of the systems.
These phenomena are attributed to an underlying
percolation transition.
In the soft potential model the transition is rounded, since each tracer encounters its own critical
density according to its energy. Therefore, the rounding of the transition is a generic occurrence in realistic, soft systems.
}


\maketitle

% footnotes must appear after maketitle
% affiliations first, use letters
\bgroup
\renewcommand\thefootnote{\alph{footnote}}
\footnotetext[1]{%
Institut f\"ur Theoretische Physik II: Weiche Materie,
Heinrich Heine-Universit\"at D\"usseldorf, Universit\"atsstra\ss{}e 1,
40225 D\"usseldorf, Germany}
\footnotetext[2]{%
Institut f\"ur Theoretische Physik, Universit\"at Erlangen--N\"urnberg, Staudtstra{\ss}e~7, 91058, Erlangen, Germany}
\footnotetext[3]{%
Max-Planck-Institut f\"ur Intelligente Systeme, Heisenbergstra{\ss}e 3, 70569 Stuttgart, Germany,
and VI. Institut f\"ur Theoretische Physik, Universit\"at Stuttgart,
Pfaffenwaldring 57, 70569 Stuttgart, Germany
}
\footnotetext[4]{
Institut f\"ur Theoretische Physik, Leopold-Franzens-Universit\"at Innsbruck, Technikerstra{\ss}e 25/2, A-6020 Innsbruck, Austria
}
\egroup

\footnotetext[2]{Correspondence: horbach@thphy.uni-duesseldorf.de}

\section{Introduction}

Molecular transport in strongly heterogeneous media is fundamental for a wide
range of disciplines such as molecular sieving~\cite{Gleiter2000}, catalysis~\cite{Brenner1993, Gleiter2000, Benichou2010, Ben-Avraham2000}
and ion-conductors \cite{Bunde1998, Voigtmann2006}, but also for protein motion in the interior of
``crowded'' cells \cite{Weiss2014, Hofling2013, Sokolov2012, Saxton2012}.
Common to all of these systems is the occurrence of a quasi-immobilized host
structure which restricts transport to ramified paths through the medium.

The generic features of transport in heterogeneous media~\cite{Adler1992, Brenner1993} can be elucidated
by studying simplified model systems such as the Lorentz model~\cite{Hofling2013}. In its simplest variant~\cite{Lorentz1905, Beijeren1982}, a point
tracer explores the space between randomly distributed hard-disk obstacles,
which may overlap and are uncorrelated.
Recently, \citet{Skinner2013} presented a colloidal realization of a
two-dimensional Lorentz model, which differs from the original model with
respect to the matrix structure and the interactions.
In detail, the experiment uses a slightly size-disparate binary mixture of
superparamagnetic colloidal spheres.
The larger particle species is immobilized by the cover slides, while the
smaller one serves as tracers.
If an external magnetic field is applied, magnetic dipoles are induced which
lead to a soft repulsion between the particles; the range of the tracer-matrix
interaction thus can be tuned by the strength of the magnetic field.

A striking observation in these models is a localization transition with
respect to the diffusive motion of the tracer particle.
In the hard-core model, long-range transport ceases to exist as a critical
obstacle density is approached~\cite{Gotze1981, Gotze1982, Hofling2006, Hofling2007,
Hofling2008, Bauer2010}.
Concomitantly, transport becomes anomalous as manifested in a non-linear,
power-law growth of the mean-squared displacement, $\msd(t) \sim t^{2/z}$
with a universal dynamic exponent $z$.
In the case of soft interactions, the experimental and simulation results~\cite{Skinner2013} indicate that the transition is rounded, i.e., the critical
singularities are seemingly avoided.
Evidence for anomalous transport has been found also in a variety of systems
with a strong dynamic asymmetry, e.g.\ in computer simulations for
alkali-doped silica melts~\cite{Horbach2002} or polymer blends with
monomer-size disparity~\cite{Moreno2007, Moreno2008}.
Recent work~\cite{Moreno2006, Moreno2006a, Voigtmann2009, Voigtmann2011} suggests that this
holds generically in size-disparate mixtures.
Although it seems plausible to expect an analogy between these more realistic models with soft interactions
and the theoretical idealization~\cite{Horbach2010}, a direct
and quantitative link is missing.

The goal of this work is to provide intermediate steps from the hard-disk
idealization to more realistic systems, thereby testing the key ingredients
leading to anomalous transport.
For hard tracer-matrix interactions, many facets of the localization
transition are well understood~\cite{Hofling2008, Spanner2013,Hofling2013}.
Most importantly, the localization transition is due to an underlying continuum
percolation transition of the accessible void space~\cite{Hofling2006}, which
is accompanied by a series of scaling laws familiar from the theory of critical
phenomena of continuous phase transitions~\cite{Ben-Avraham2000}.
Above a critical obstacle density, the network of the void space falls apart
into a hierarchy of finite-sized pores.
At criticality, the void space is a self-similar fractal in the statistical
sense, which entails subdiffusion for tracers exploring these structures~\cite{Ben-Avraham2000, Kammerer2008}.
Correlations in the host matrix modify the geometry of the void space with
potential implications on the critical behavior.
Moreover, soft interactions smear out the boundaries of the accessible space
and change the topology of the percolation network by introducing a potential
energy landscape with finite barriers between the pores.

To investigate these issues, we compare simulations of two different models, which represent modifications to the original Lorentz model. We focus on two-dimensional systems which are amenable to colloid experiments.
First, we introduce spatial correlations in the host matrix by using an
extended tracer in frozen-in configurations of equilibrated hard disks.
The resulting host structure is equivalent to the cherry-pit model
\cite{Torquato2002}.
Second, we relax the assumption of hard-core repulsion between both obstacles
and tracers by introducing soft interactions.
Now, the host structures are generated from snapshots of an equilibrated fluid
of soft particles.
By this, the transition becomes rounded and we demonstrate that the rounding
originates naturally from the energy distribution of the tracers.
We find that an effective interaction distance can be assigned to each tracer
according to its energy, thereby providing a mapping to the hard-core case.

\begin{figure}
		\includegraphics[width=\columnwidth]{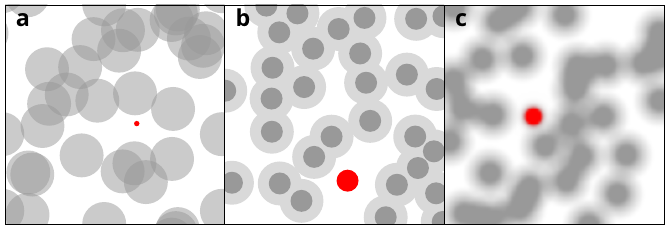}
		\includegraphics[width=\columnwidth]{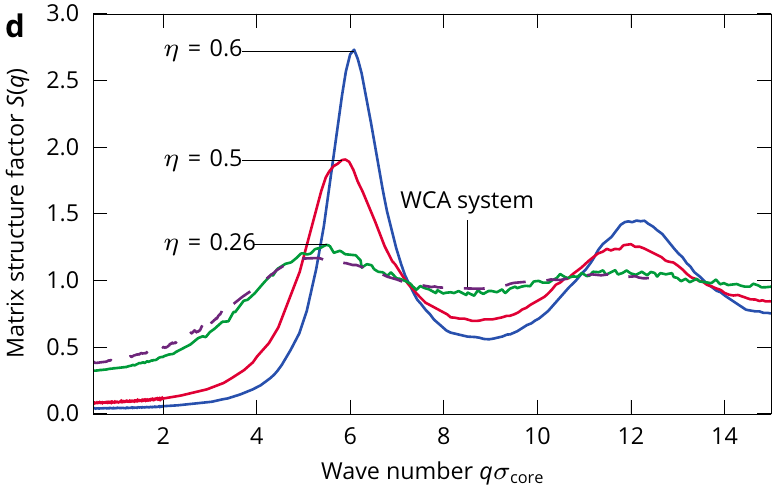}

\caption{a-c) Illustrations of the relevant models. a) Lorentz model: overlapping obstacles (grey) and point tracer (red). b) Cherry-pit model: obstacles (dark grey) and extended tracer (red). The area inaccessible to the tracer center is marked in light grey. c) WCA-system: soft obstacles (grey) and soft tracer (red). 
d) Static structure factor of the matrix in the cherry-pit model.
For comparison, the matrix structure factor of the 
WCA-disk system with the effective diameter $\sigma\core\BH$ is included. } 
\label{fig:sofq}
\end{figure}

\section{Cherry-pit model}
\label{sect:cherryPit}

\paragraph{Host structures}

In the cherry-pit model, the matrix of obstacles is formed by equilibrium
configurations of $N$ non-overlapping disks of diameter $\sigma\core$,
packing fraction $\eta = (N/L^2) \pi \sigma\core^2/4$,
and the centers are confined to a square of edge length~$L$.
The remaining space is explored by a ``ballistic'' tracer undergoing
specular scattering from the obstacles, yielding trajectories $\vec R(t)$ which
are piece-wise straight lines.
The tracer particles are disks of finite diameter~$\sigma_\text{T}$, contrarily to the original overlapping
Lorentz model (\cref{fig:sofq}a,b).
For comparison to the latter, we introduce the interaction
distance $\sigma := (\sigma\core + \sigma_\text{T})/2$, with which a dimensionless control parameter, the
reduced number density $n\reduced:= (N/L^2) \sigma^2$, can be defined.
The velocity of the tracer is of fixed magnitude $\vel$
and defines a time scale
$t_o=\sigma/\vel$. Transport is controlled by variation of $n\reduced$ at fixed $\eta$.

The matrix configurations are generated by canonical Monte Carlo simulations, where
the particles are initially placed onto a hexagonal grid.
We consider systems of $N=10{,}044$ or $N=516{,}468$ disks in a square
box with varying size assuming periodic boundary conditions in
the two spatial directions.
For the equilibration of the systems we combine displacement moves with
cluster moves proposed by \citet{Dress1995}.
In the displacement moves, a random particle $i$ with position $\vec r_i$ is 
displaced to a new position $\vec r_i+ \vec{\delta}$, where the vector 
$\vec{\delta}$ is randomly chosen such that $|\vec{\delta}| < \sigma\core$.
This move is accepted according to a standard Metropolis criterion\cite{Metropolis1953}.

Cluster moves are applied periodically after 10 displacement moves.
To this end, a pivot is selected as a random point in the system.
By starting with one randomly selected disk and recursively searching for disks
overlapping with the disks' mirror image with respect to the pivot, we identify
a pair of disk clusters ($C_1,C_2)$, $C_1\ne C_2$, defined as two sets of
disks satisfying the following condition:
When all disks in $C_1$ are reflected at the pivot, each of them overlaps with at least one disk in $C_2$, but none overlaps with disks not in $C_2$, and vice versa.
If clusters are larger than 15 disks, the cluster move is rejected.
In this manner, the clusters can be exchanged with their reflected counterparts.
In the following, the Monte Carlo time is given in terms of \emph{cycles},
where each cycle consists of $N$ displacement moves and $N/10$ cluster moves.

Systems with packing fractions ranging
from $\eta=0.02$ to $0.90$ for the small systems and $\eta=0.02$ to $0.65$ for
the large systems were generated.
At each value of $\eta$, the configurations were first equilibrated for at least 1,000 cycles
for low packing fractions and up to 50,000 cycles for high packing fractions
to ensure proper equilibration, particularly for $\eta \lesssim 0.7$, i.e.~for packing fractions lower than the location of the fluid-to-solid transition in hard disks \cite{Bernard2011, Kapfer2014}.
To check whether the system was sufficiently equilibrated we monitored the structure factor $S(q)$ and the pair correlation function $g(r)$
and compared it to the Percus--Yevick approximation~\cite{Percus1958,Adda-Bedia2008}. Additionally, for systems $\eta < 0.7$, we required that particles are displaced by $L/2$ on average.
For each equilibrated configuration, a production run was performed to yield 20 independent configurations, each of them separated by the respective equilibration time. These configurations served as matrix configurations for the tracer particle dynamics.

The structural correlations contained in the obstacle matrix are a function of the packing fraction $\eta$ and are measured by the static structure factor of the obstacles,
\begin{align}
	S(q) =
	\frac{1}{N} \av{ \sum_{j,k=1}^N \exp[ -i \vec q \cdot (\vec {r}_j - \vec {r}_k)]},
\end{align}
as a function of the wave number $q = |\vec q|$,  see~\cref{fig:sofq}. The angled brackets represent an ensemble average over the disorder, and $\{ \vec r_j \}$ denote the positions of obstacle centers, $j=1,\dots,N$. At low packing fractions the system exhibits the structure of a dilute liquid, as indicated by the low amplitude of the first diffraction peak, e.g. $S(q_\text{max})\approx 1.3$ at $\eta = 0.26$. As the packing fractions increases, the peak grows in amplitude, e.g. $S(q_\text{max})\approx 2.7$ at $\eta = 0.6$, and $S(q)$ exhibits pronounced short-range order, indicating the structure of a dense liquid.

\begin{figure}
\centering
	\includegraphics[width = \columnwidth]{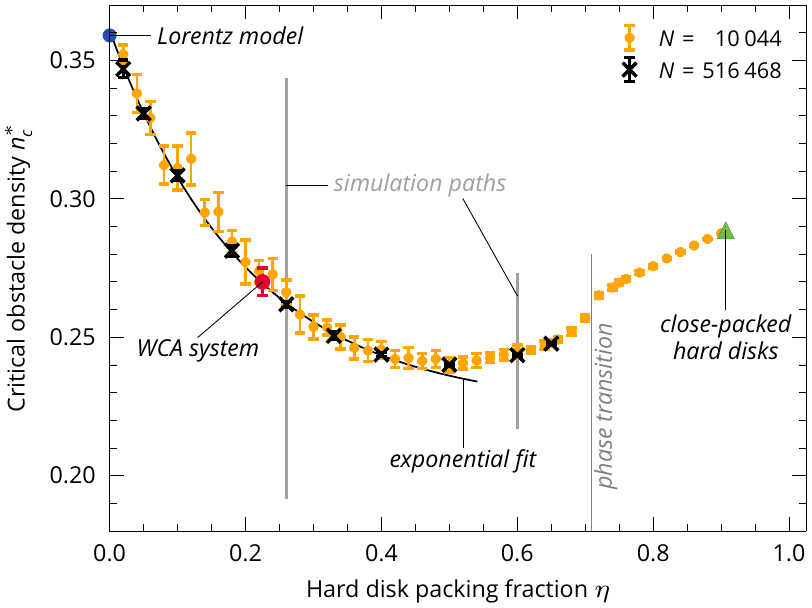}
    \caption{Critical reduced density $n\reduced\crit$ for the
    cherry-pit model as a function of the packing fraction $\eta$ of the
    obstacle cores for two different system sizes, $10^4$ and $\approx 5\cdot
    10^5$ obstacles.  The overlapping Lorentz model corresponds to $\eta = 0$. For
    comparison: WCA system at the effective packing fraction 
		$\eta = 0.225$ and $n\reduced\crit
    \approx 0.272$.
		}
	\label{fig:percolationThreshold}
\end{figure}

\paragraph{Percolation threshold}

For the study of the localization transition, it is crucial to precisely know the percolation threshold of the
void space accessible to the tracer particle.
For a given obstacle configuration, we have determined the threshold value
$n\reduced\crit=N \sigma\crit^2/L^2$ of the reduced number density
by varying the distance~$\sigma$ of the tracer--obstacle interaction.
First, we have computed the edges of a Voronoi tesselation
of the matrix using the free \texttt{voro++} software library~\cite{Rycroft2009}.
After removal of the edges with a distance smaller than $\sigma$
to an obstacle center, the obtained network represents the volume accessible to the tracer particle~\cite{Kerstein1983}.
Upon increasing $\sigma$, this connectivity network is diluted
until the critical value  $\sigma\crit$ is reached, where the residual network
barely spans the entire simulation box.

While there is a unique critical density $n\reduced\crit$ for infinitely large systems $L\to\infty$, at finite system sizes $L$ the percolation thresholds of the individual obstacle configurations follow a distribution with a finite width. For decreasing system size, the mean of the distribution is shifted towards a slightly higher critical density $n\reduced\crit(L)$ according to $n\reduced\crit(L)-n\reduced\crit \sim L^{-1/\nu}$ \cite{Hofling2008}.
Additionally, the width of the distribution, which can be measured with the standard deviation $\delta n\reduced\crit(L)$ for example, scales as $\delta n\reduced\crit(L)\sim L^{-1/\nu}$.

Over the full range of packing fractions $0<\eta<\eta_\text{hcp}$ from the ideal gas to close packing,
we have numerically determined the
critical reduced density $n\reduced\crit(\eta)$ for two different system sizes, shown in \cref{fig:percolationThreshold}. The critical density is calculated from the mean of the percolation distance $\sigma\crit$ of the obstacle configurations. The error bars are calculated with the help of the relative standard deviation of the critical distance $\Delta(\eta) := \delta\sigma\crit(\eta)/\sigma\crit(\eta)$ and thus give an estimate of the width of the distribution of the percolation thresholds. This gives an estimate for the percolation density, $n\reduced\crit(\eta) = (N \sigma\crit^2(\eta)/L^2)\cdot[1\pm \Delta(\eta)]^2$. 

We confirmed exemplarily for the case $\eta=0.26$ that the relative standard deviation $\Delta(\eta)$ is indeed a good approximation to the distribution width $\delta n\reduced\crit(\eta,L)$ of the critical distance, as we observed that increasing the number of independent configurations up to 300 did not modify $\Delta(\eta)$ within the specified precision.

The overlapping Lorentz model corresponds to $\eta=0$, here the percolation threshold is known accurately~\cite{Quintanilla2007}: $n\reduced\crit(0) = 0.359\,081\,0 \pm 0.000\,000\,6$
for the infinitely large system.
For packing fractions $\eta \lesssim 0.45$, the
percolation threshold decreases from this value and can be fitted with
a shifted exponential function $f(\eta)= a \exp(-b\eta)+c$. 
For $0.45 \lesssim \eta \lesssim 0.9$, the percolation threshold is growing, with a ``shoulder'' around $\eta \approx 0.7$ indicating the 2D melting transition.
At $\eta_\text{hcp} = (\pi/6)\sqrt{3} \approx 0.9$, the system displays a hexagonal closed-packed structure and therefore $n\reduced\crit = \sqrt{3}/6 \approx 0.289$.

For the following study of the tracer dynamics and how it is affected by the structural correlations contained in the matrix,
we consider in detail $\eta=0.26$ and $\eta=0.60$ with percolation thresholds
$n\reduced\crit = 0.262 \cdot (1\pm 2\cdot 10^{-3})^2$ and $n\reduced\crit = 0.2442 \cdot (1\pm 7\cdot 10^{-4})^2$ respectively.
Note that at $\eta=0.26$ the structure factor closely resembles that of the WCA system discussed later on (\cref{fig:sofq}d).

\begin{figure}
	\centering
 	\includegraphics[width=0.95\columnwidth]{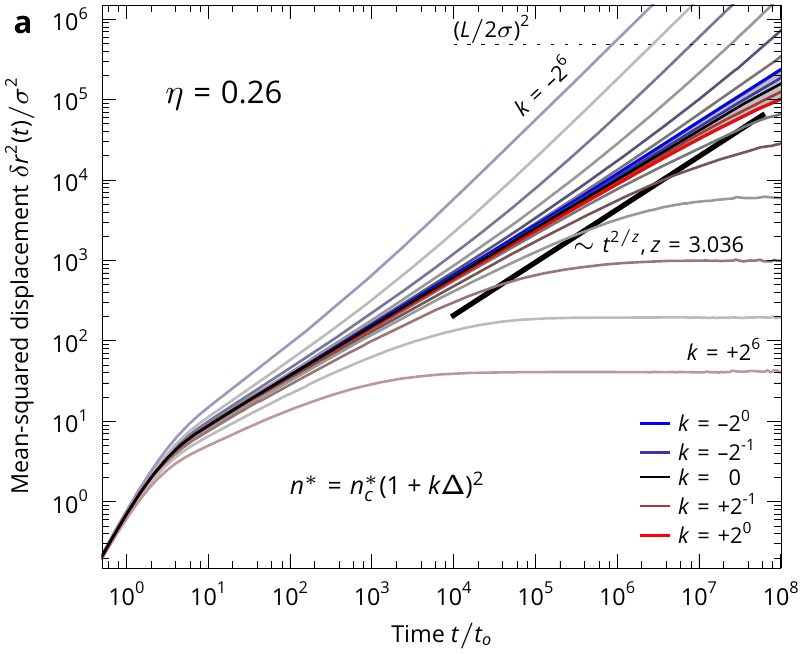}
	\includegraphics[width=0.95\columnwidth]{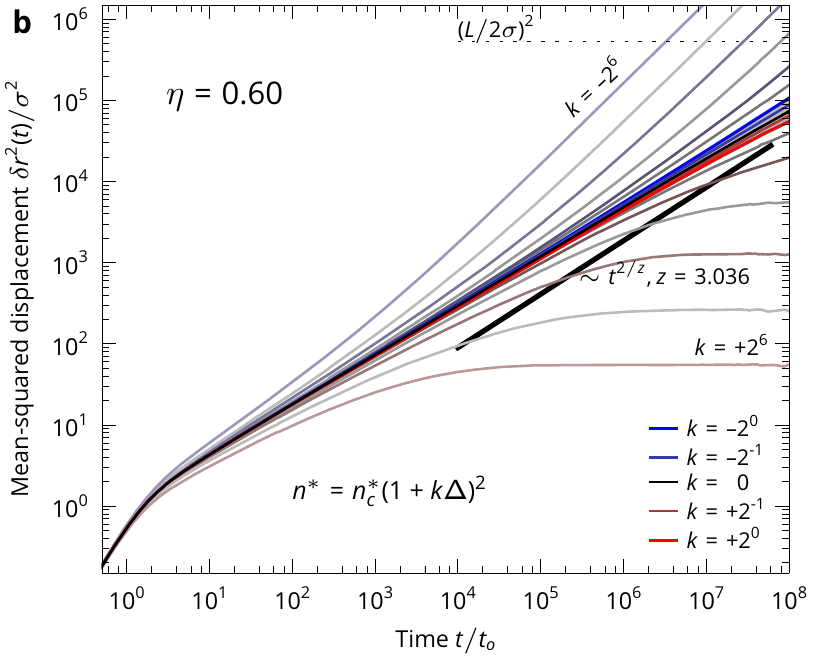}
    \caption{Mean-squared displacements in the cherry-pit model. 
Reduced    density $n\reduced$ is varied around the critical density $n\reduced=  n\reduced\crit (1+k \Delta)^2$, $k= 0, \pm 2^{-1}, \ldots \pm 2^{6}$
in geometric progression.  Obstacle packing fractions (a) $\eta = 0.26$, $n\crit\reduced= 0.262$ with standard deviation of the percolation threshold $\Delta= 2\cdot 10^{-3}$ and (b) $\eta = 0.6$, $n\crit\reduced =0.2442 $
and $\Delta = 7\cdot 10^{-4}$. 
Data below $n\reduced\crit$  fan out towards the upper left, the ones above $n\reduced\crit$ towards the lower right.  The solid
    line indicates a power-law $\propto t^{2/z}$ with the dynamic exponent of
    the Lorentz model $z = 3.036$. The horizontal dashed line indicates the size of the simulation box.
}
    \label{fig:msd_cp}
\end{figure}

\paragraph{Tracer dynamics}

The mean-squared displacements (MSD) $\msd(t) := \av{|\vec R(t) - \vec R(0)|^2}$ were obtained as time- and ensemble-average over 20 obstacle configurations containing $N=516,468$ obstacles each in production runs up to times  $10^9t_o$. Each obstacle configuration was probed by at least 8 tracers, while 32 tracers were used close to~$n\reduced\crit$. For each tracer a random point in the void space was chosen as the initial position. Moving time averages were calculated efficiently with an ``order-n'' algorithm~\cite{Frenkel2002, Colberg2011}.

The MSD are shown in \cref{fig:msd_cp}
  for reduced densities close to the critical one such that the interaction distance $\sigma$ is changed in geometric progression with the standard deviation $\Delta$ as basic scale.   
For both values of $\eta$, the localization transition is
evident and qualitatively similar to the overlapping 2D Lorentz model
% (\cref{fig:msd_LM})
\cite{Bauer2010, Hofling2007, Hofling2014}:
For times $t$ longer than a certain crossover time scale $t_x$, 
the MSD either grows 
diffusively, $\msd(t) \simeq 4 D t$ for $n\reduced < n\reduced\crit$ with diffusion coefficient $D$, or 
saturates, $\msd(t) \simeq \ell^2$ for $n\reduced > n\reduced\crit$ with localization length~$\ell$.
The transport is highly heterogeneous in space:
a fraction of tracers is confined to finite pores, which exist at all densities and have a broad distribution of sizes near $n\reduced\crit$,
but only tracers on the spanning cluster contribute to long-range transport.
For $n\reduced > n\reduced\crit$, the spanning cluster disappears and \emph{all} tracers are
confined. This implies that the localization length $\ell$ is the root-mean-square size of the finite clusters.
As the critical point is approached, $n\reduced \to n\reduced\crit$, a
\emph{sub}-diffusive regime emerges in a growing time window,
\begin{equation}
  \msd(t)\sim t^{2/z} \,, \quad t_o \ll t \ll t_x \,.
\end{equation}
The exponent $z$ is believed to be universal for particle
transport on 2D percolation clusters~\cite{Halperin1985,
Machta1986}
and may be considered the fundamental dynamic exponent of the problem.
It was estimated to $z=3.036 \pm 0.001$
from studies of the conductivity of random resistor networks~\cite{Grassberger1999},
random walkers on percolation lattices~\cite{Kammerer2008}
and in the overlapping 2D Lorentz model~\cite{Bauer2010, Hofling2011}.
The value was confirmed only recently also for the overlapping 2D Lorentz model with ballistic tracers~\cite{Hofling2014}.
Our data for the  MSD in the cherry-pit model suggest anomalous transport with effective exponents slightly lower than the universal one (\cref{fig:msd_cp}).

\begin{figure}
	\centering
 \includegraphics[width=0.95\columnwidth]{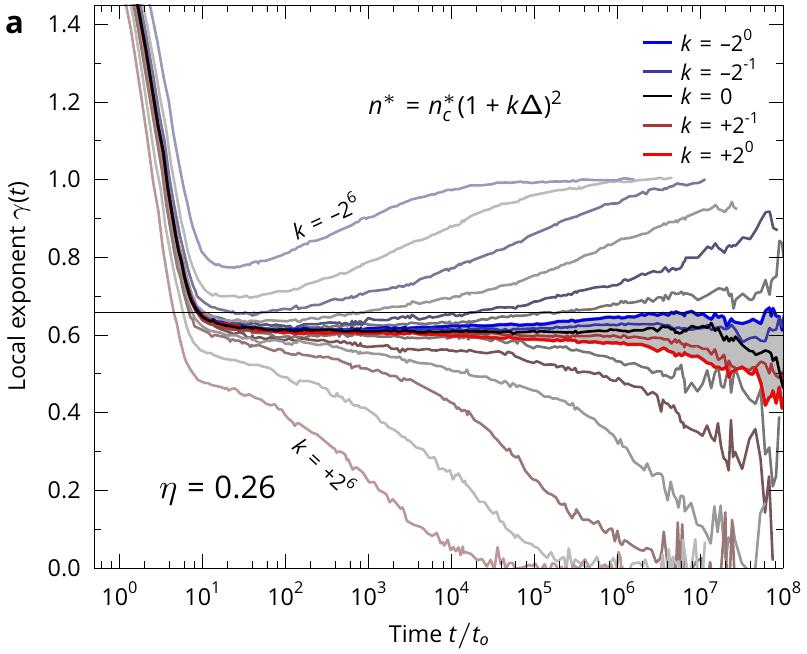}
 \includegraphics[width=0.95\columnwidth]{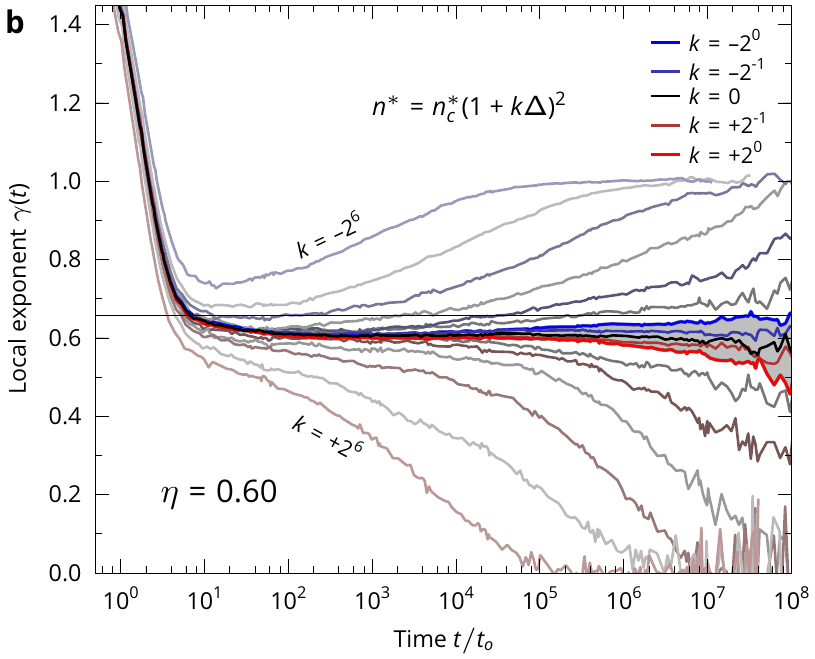}
    \caption{Local exponents of the mean-squared displacements of
    \cref{fig:msd_cp} of the cherry-pit model for the obstacle packing
    fractions (a) $\eta = 0.26$ and (b) $\eta = 0.6$. Reduced density increases
    from top to bottom. The horizontal line indicates the anomalous exponent
    $2/z$ with $z = 3.036$ of the Lorentz model. The shaded areas correspond to one standard deviation $\Delta$ in the interaction distance $\sigma$. 
}
    \label{fig:gamma_cp}
\end{figure}

A more thorough test of the value of the anomalous exponent can be achieved with
the local exponent
$\gamma(t)$ of the MSD defined as
\begin{align}
    \gamma(t) := \frac{\diff \log\boldsymbol(\msd(t)\boldsymbol)}{\diff \log(t)} \,.
\end{align}
At short times, the exponent $\gamma(t)$ decays quickly from its initial value
2 for ballistic motion due to the scattering from the obstacles. 
At the lowest densities, $\gamma(t)$ rapidly converges to 1 corresponding to the  linear increase of the MSD. At high densities the local exponents converge to 0, reflecting the localization. Values corresponding to anomalous diffusion are found close to the transition, yet the exponent found here seems to underestimate the universal value, obeyed by a random walker on percolation lattices~\cite{Kammerer2008}. 
However, the dynamics is extremely sensitive to the density near the percolation threshold. 
Also, the local exponent becomes compatible with the universal value of $z$ at the lower end of the error margin for the percolation threshold, see blue lines for $k=-2^0$ in \cref{fig:gamma_cp}.
Additionally, the overlapping 2D Lorentz model with ballistic dynamics exhibits strong, non-universal corrections to scaling, which modify the effective exponent over long periods of time.~\cite{Hofling2014} In particular, $\gamma(t)$ slowly approaches $2/z$ from below. It is thus entirely expected that the cherry-pit exhibits a similarly slow convergence. 

\begin{figure}
	\centering
		\includegraphics[width=\columnwidth]{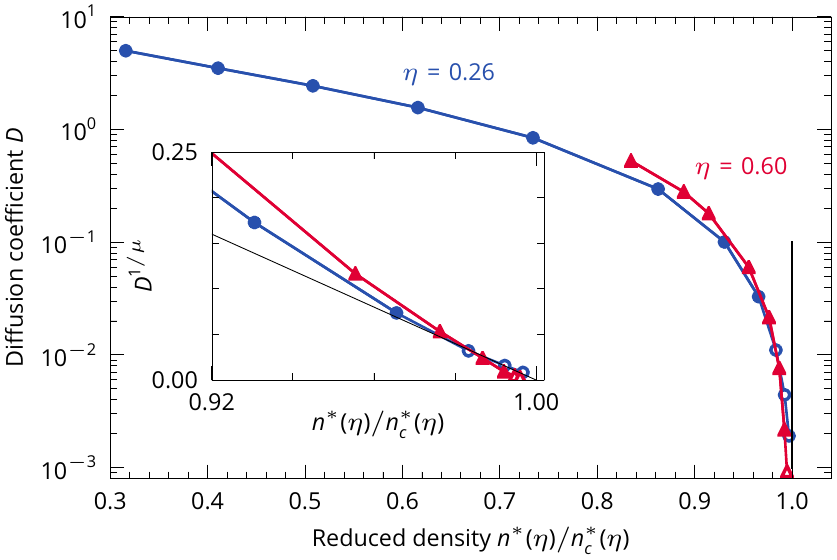}
    \caption{Diffusion coefficients for the cherry-pit model  at packing fractions $\eta = 0.26$ and $0.6$ as a function of
    the reduced density $n\reduced$ divided by the respective percolation thresholds $n\reduced\crit(\eta) = 0.262$ and $0.2442$, respectively. Open symbols mark data points which were obtained at densities where the MSD had not quite become diffusive and thus potentially overestimate $D$. 
    Inset: Rectification plot of the same data testing \cref{eq:D_scaling} with conductivity exponent $\mu= 1.309$.
    The straight line serves as guide to the eye.
  }
    \label{fig:D_cp_n0p26}
\end{figure}

In the approach to the percolation threshold, long-time diffusion decreases such that it vanishes at the critical point.
Scaling arguments~\cite{Ben-Avraham2000, Gefen1983} predict a power-law singularity,
\begin{equation}
  D(\epsilon \uparrow 0) \sim (-\epsilon)^\mu \,, \quad
  \epsilon := (n\reduced - n\reduced\crit) / n\reduced\crit \,,
  \label{eq:D_scaling}
\end{equation}
with the exponent fixed by $\mu=(z - 2)(\nu - \beta/2)$.
The universal exponents $\nu$ and $\beta$ characterize the
underlying geometry of the spanning cluster, namely its correlation length $\xi \sim
|\epsilon|^{-\nu}$ (the scale up to which it is self-similar) and its weight
$P_\infty \sim |\epsilon|^\beta$.
For two-dimensional standard percolation, $\nu=4/3$ and $\beta=5/36$ hold exactly
\cite{Ben-Avraham2000}, and one computes $\mu=1.309\pm 0.002$ from the above value of~$z$.

The diffusion coefficients obtained from the long-time behavior of the MSDs are reduced for larger tracers (at fixed $\eta$, $\sigma\core$),
and the suppression of diffusion is compatible with the percolation threshold as determined above,
see \cref{fig:D_cp_n0p26}. 
Plotting $D^{1/\mu}$ \emph{vs.} $n\reduced$ (see inset) rectifies the critical law, \cref{eq:D_scaling},
and would yield a straight line if the power law was an accurate description over the full range $n\reduced<n\reduced\crit$ .
From the data one infers that the scaling law becomes valid for $\epsilon \lesssim 0.05$, which is similar to the situation in the overlapping case~\cite{Hofling2014}.

\section{WCA-disk systems}

\paragraph{Host structure}

Next, we move towards possible experimental realizations and relax the idealization of a hard-core exclusion between the particles, considering soft interactions.
To serve as frozen host structures, we take snapshots of an equilibrated liquid of polydisperse particles at moderate densities. The particles interact via a Weeks--Chandler--Andersen (WCA) potential\cite{Weeks1971}, which is a truncated and shifted Lennard-Jones potential such that the interaction is purely repulsive,
\begin{align} 
V_{\alpha\beta}(r) = \begin{cases}
4\eps\matrx \left[
\left(\frac{\sigma_{\alpha\beta}}{r}\right)^{12}  -
\left(\frac{\sigma_{\alpha\beta}}{r}\right)^{6}  + \frac{1}{4}\right] 
, & r<r\cut, \\ 
0, & r\geq r\cut, \end{cases} \label{eq:WCA} 
\end{align}
with a cutoff $r\cut := 2^{1/6} \sigma_{\alpha\beta}$.
To avoid crystallization, a polydisperse mixture is necessary. To this end, the diameters of the $N$ particles are chosen to be additive, $\sigma_{\alpha\beta}:=
(\sigma_\alpha+\sigma_\beta)/2$, and are taken equidistantly from an interval,
$\sigma_\alpha = (0.85 + 0.3\,\alpha/N)\sigma\matrx$ with $\alpha,\beta = 1,\ldots,N$. The units of length and energy are fixed by $\sigma\matrx$ and $\eps\matrx$, respectively. The temperature is set to
$k_B T/\eps\matrx = 1.0$. 
To improve numerical stability the potential is multiplied with a smoothing function $\Psi(r) := (r-r\cut)^4/[h^4+(r-r\cut)^4]$ with the width $h=0.005\, \sigma\matrx$.

The particle configurations are equilibrated using a simplified Andersen thermostat
\cite{Andersen1980} by randomly drawing their velocities from a Maxwell
distribution every 100 steps with thermal velocity $\text{v}\th := (k_BT/m)^{1/2}$. We use the Lennard-Jones time $t_o := \sigma\matrx/\text{v}\th = [m(\sigma\matrx)^2/\eps\matrx]^{1/2}$ as basic unit of time. Newton's equations of motion are integrated numerically with the velocity-Verlet algorithm\cite{Binder2004} using a numerical timestep of $\Delta t = 7.2\cdot 10^{-4}t_o$.

We generated 100 statistically independent host structures 
for particle numbers $N = 500$, $1\,000$, $2\,000$, $4\,000$, and $16\,000$
at fixed number density $n := N/L^2 =  0.278\,(\sigma\matrx)^{-2}$, corresponding to
system sizes $L/\sigma\matrx = 42.4$, $60$, $84.8$, $120$, and $240$.

With each generated structure we associate a percolation threshold relying on a Voronoi tesselation of the particle positions of the host structure, in the same way as for the cherry-pit systems, see \cref{sect:cherryPit}. Averaging over all 100 snapshots at the largest system size yields a critical effective interaction distance $\sigma\crit/\sigma\matrx = 0.990\pm 0.009$ or equivalently, a critical reduced obstacle density  $n\reduced\crit  := n\sigma\crit^2  = 0.272\pm 0.005$.

It is instructive to structurally compare the WCA system to the cherry-pit model, employing an effective hard-core diameter. Here we use the Barker--Henderson diameter $\sigma\core\BH$, originally developed in the context of thermodynamic perturbation theory\cite{Barker1967, Henderson1977, Hansen2006},
\begin{align}
	\sigma\core\BH = \int_o^\infty (1 - e^{-\beta V_{\alpha\beta}(r)})\ \diff r.
\end{align}
Numerical evaluation of the integral for $\sigma_{\alpha\beta} = \sigma\matrx$ yields $\sigma\core\BH \approx 1.02\, \sigma\matrx$, corresponding to an effective packing fraction of $\eta := n \pi (\sigma\core\BH)^2/4 = 0.225$. At this packing fraction, the cherry-pit model exhibits a similar percolation threshold, see \cref{fig:percolationThreshold}.

The structure factor of the WCA system is included in \cref{fig:sofq}, with the wave numbers measured in units of $1/\sigma\core\BH$. It compares well to the one of the slightly denser cherry-pit system at $\eta = 0.26$. The positions of the first diffraction peak coincide,  while the amplitude in the WCA system is slightly lower by $\approx 9\%$.
Thus, the percolation threshold, the effective reduced density, and the structure factor of the WCA-system matrix can be mapped consistently onto the cherry-pit model.

The frozen matrices are explored by an ideal gas of tracers. The tracers interact with all matrix particles identically via the smoothly truncated WCA potential, \cref{eq:WCA}, with coefficients $\eps\tracer := 0.1 \eps\matrx$ and $\sigma_{\alpha\beta} := \sigma\tracer$.
The interaction range $\sigma\tracer$ is used as the control parameter and defines a reduced number density $n\WCA\reduced:= n (\sigma\tracer)^2$. In the experiment by \citet{Skinner2013}, the tuning of the analogous tracer--matrix interaction is achieved by varying the external magnetic field.
The tracer particles are inserted and equilibrated in the host structure by grand-canonical Monte-Carlo moves in combination with successive umbrella sampling.~\cite{Virnau2004}
Subsequently, the tracers are equilibrated using the simplified Andersen thermostat.  
Since the equilibration is performed in the canonical ensemble the average energy of each system is fluctuating. For the micronanonical production runs the systems are brought to the same average energy at the end of the equilibration period
by rescaling all tracer velocities in the same system with the
same constant, leaving the relative distribution of energies unchanged.
 
Newton's equations of motion are integrated numerically with the velocity-Verlet algorithm with the same timestep as for the host particles.
Between 50 and 10,000 tracers for each host structure configuration are used to obtain ensemble averages for runs of up to nearly $10^6 t_o$. For the calculation of time averages, typically 10 moving time origins per run were used and were spaced equidistantly over the whole simulation time.

 \begin{figure} [tb]
 	\centering
 		\includegraphics{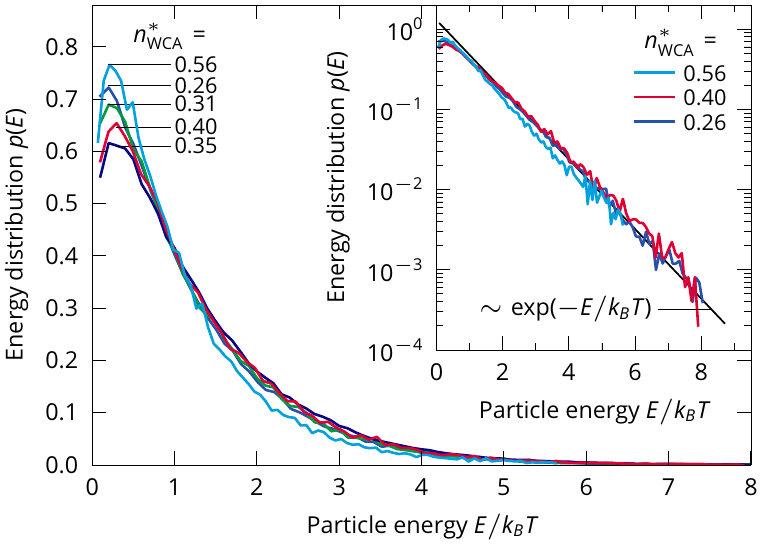}
    \caption{Energy distribution $p(E)$ in the WCA-disk system for a canonical ensemble of tracers for a range of reduced densities $n\reduced$.
		 Inset: the same data in semilogarithmic presentation.}
    \label{fig:energy_histogram_es}
 \end{figure}

The probability distribution of the energy per tracer $p(E)$, as defined by $p(E) = Z(\beta)^{-1} \exp[-\beta E] \mathcal D(E)$ with the density of states $\mathcal D(E)$ and the partition function $Z(\beta)$, can be directly calculated from the simulation data as the histogram of the tracer energy. For the binning of the energies, a bin width of $\Delta E/\eps\matrx = 0.1$ was chosen. The distribution $p(E)$ has a peak at small energies, see \cref{fig:energy_histogram_es}, and decays exponentially at large energies, see inset. The energy distribution is nearly unchanged for all densities $n\WCA\reduced$. Merely slight variations in the peak height are observed, which are probably due to fluctuations in the potential energy frozen into the matrix.

\begin{figure}
	\centering
		\includegraphics[width=\columnwidth]{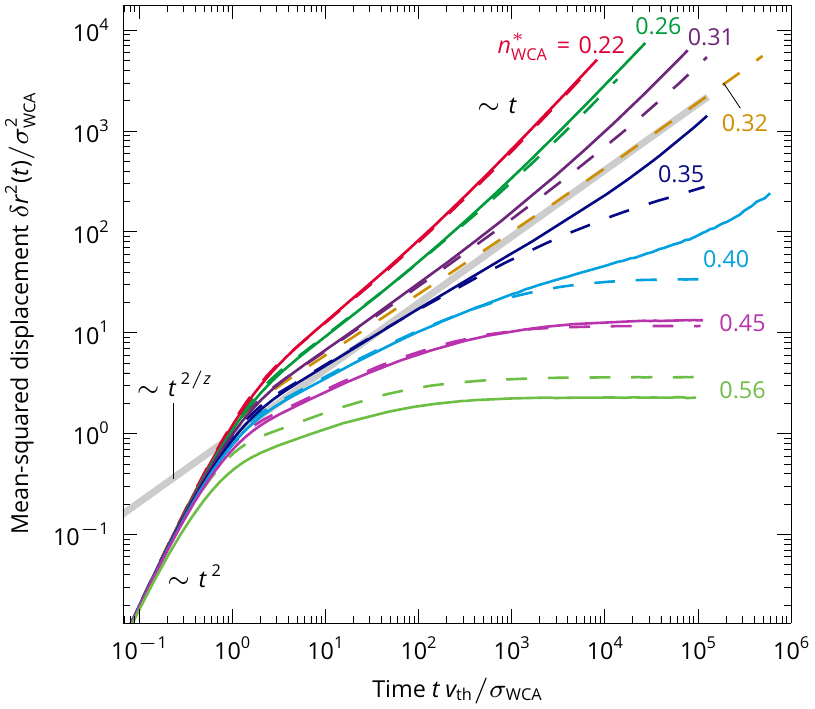}
    \caption{Mean-squared displacements of the WCA system for a canonical 
		ensemble of tracers
    (solid lines) and tracers with exactly one energy (dashed lines) for a  
    range of $n\reduced\WCA$.
    The straight line $\sim t^{2/z}$ with the dynamic exponent $z =
    3.036$ of the Lorentz model serves as guide to the eye. (Data published previously in Ref.~\citenum{Skinner2013})
				 }
    \label{fig:WCA_MSD}
\end{figure}

\begin{figure}
	\centering
	\includegraphics[width=\columnwidth]{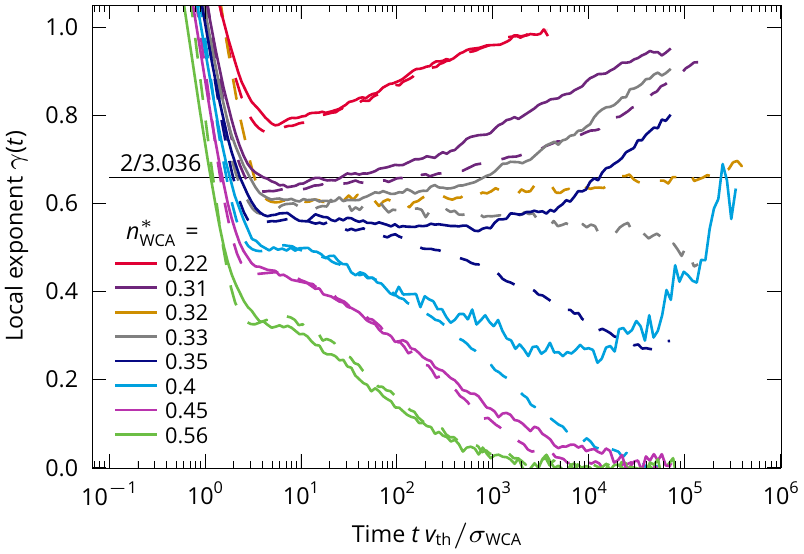}
    \caption{Local exponent of the mean-squared displacements of the WCA system  for the same data as in \cref{fig:WCA_MSD}, i.e. for a canonical ensemble of tracers (solid lines) and tracers with exactly
		 one energy (dashed lines). The
		 horizontal line
    indicates the anomalous exponent $2/z$ with $z = 3.036$ of the Lorentz
    model.
		 }
    \label{fig:WCA_effectiveExponent}
\end{figure}

\paragraph{Tracer dynamics}

The system undergoes a localization transition similarly to the overlapping Lorentz model and the cherry-pit model: At long times, the MSD becomes diffusive for $n\WCA\reduced \leq 0.4$ and saturates for $n\WCA\reduced>0.4$, see solid lines in \cref{fig:WCA_MSD}. This implies a transition point $(n\WCA\reduced)\crit \approx 0.4$. At intermediate densities, the MSD is subdiffusive at intermediate times but it never matches the critical subdiffusion of the Lorentz model, with  exponent $2/z$ with $z = 3.036$.
This was already discussed shortly by some of us in Ref.~\citenum{Skinner2013}. This is even more apparent by direct inspection of 
the local exponent $\gamma(t)$, see solid lines in \cref{fig:WCA_effectiveExponent}. 
Instead of the Lorentz model exponent, the local exponent exhibits $\gamma(t) \approx 0.55$ at $n\WCA\reduced=0.35$ over almost three orders of magnitude in time.

The situation changes qualitatively if all tracers are set to exactly the same energy, which restores the critical behavior\cite{Skinner2013}. Then, the system undergoes a localization transition at $n\WCA\reduced \approx 0.320$, where the MSD exhibits subdiffusion with the expected exponent, see dashed lines in \cref{fig:WCA_MSD} and \cref{fig:WCA_effectiveExponent}.

\begin{figure}
	\includegraphics{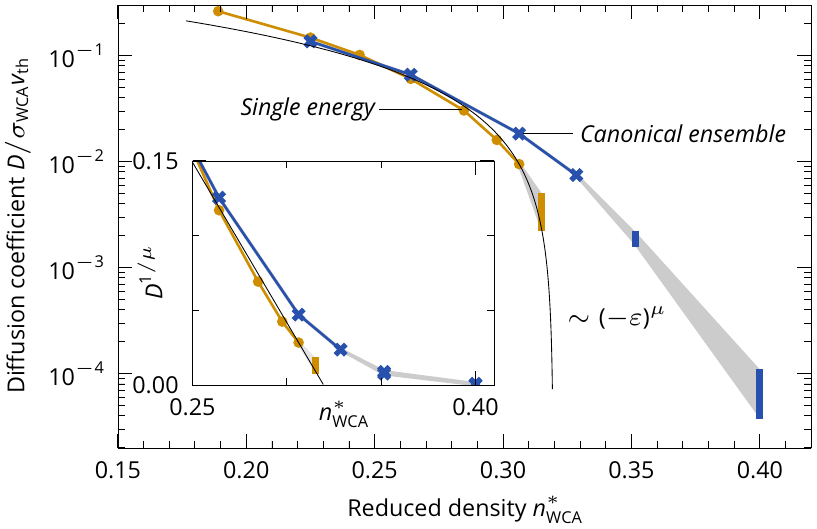}
    \caption{Diffusion coefficient $D$ of the WCA system for a canonical ensemble of tracers and the single-energy case as function of the reduced obstacle 
		density $n\reduced\WCA$. Connected symbols are obtained directly from
    the mean-squared displacements, isolated errorbars at higher densities from
    finite-size scaling, see text. The solid black line $\propto (-\eps)^{\mu}$
    with the conductivity exponent $\mu = 1.309$ of the Lorentz model serves as
    guide to the eye. Inset: Rectification plot of
    the same data.
    } \label{fig:WCA_D}
\end{figure}

\begin{figure}
	\centering
	\includegraphics{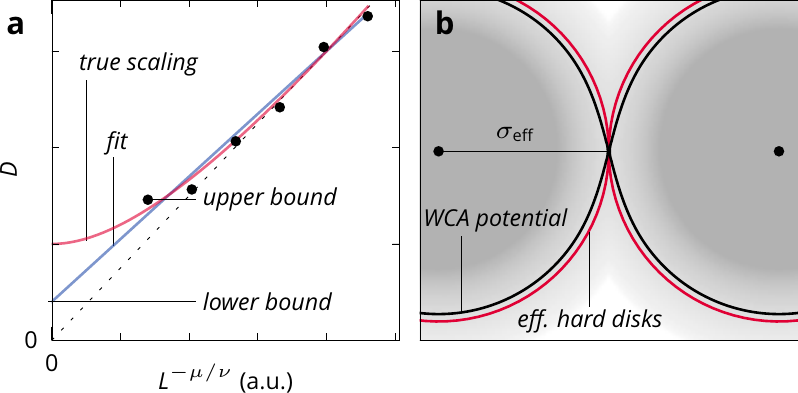}
	\caption{a) Schematic representation of the finite size scaling of the diffusion coefficient $D$ at some finite but small distance $\epsilon$ to the localization transition. Dots represent data obtained from simulations (not actually simulated here) plotted as a function of $L^{-\mu/\nu}$. The true scaling (red line) is some unknown function fulfilling $L^{-\mu/\nu}$ at small $L$. Fitting to this small-$L$ part provides a lower bound for $D$ as $L\to \infty$.
	b) Illustration of a channel between two obstacles at distance $2\sigma\eff$ with the potential energy in greyscale. Obstacle centers are marked by dots, the equipotential line of the WCA potential at the energy $E = 2V_{\alpha\beta}(\sigma\eff)$ where the channel closes is given in black, and the corresponding effective hard disks are given by red circles.}
	\label{fig:WCA_D_schematic}
\end{figure}

The difference between these two systems is also strikingly apparent in the long-time diffusion coefficient, shown in \cref{fig:WCA_D}. 
At large densities, the diffusion coefficient could not be directly measured from the MSD but was determined via finite-size scaling. 
For small separations $\epsilon$ from the critical point, the diffusion coefficient $D$ is expected to vanish as $D \sim (-\epsilon)^\mu$ for $\epsilon \to 0$. If the size of the simulation box $L$ is smaller than the correlation length $\xi$ of the system, then this scaling is replaced by the finite-size scaling $D \sim L^{-\mu/\nu}$ for $L \ll \xi$ \cite{Hofling2008}.
For constant $\epsilon$ and incrementally increasing $L$, the diffusion coefficient will first follow the finite-size scaling $D \sim L^{-\mu/\nu}$ before converging to the true value at large-enough $L\gg\xi$.
This behavior is approximately fulfilled by the fitting function $D = a L^{-\mu/\nu} + D_{\text{lower}}$. Therefore, even if the simulated systems are not large enough to allow determining the true value of $D$, a fit to the small-$L$ data will return a true lower bound $D_{\text{lower}}$, see \cref{fig:WCA_D_schematic} for an illustration. A true upper bound for $D$ is given by the value obtained in the largest simulated system. With this procedure we calculated the bounds shown as vertical bars in \cref{fig:WCA_D}.

While the data of the canonical ensemble of tracers is not compatible with the
critical behavior of the Lorentz model, $D\sim(-\epsilon)^\mu$, the single-energy case is.
The difference between the two cases is even starker in the rectification plot given in the inset of \cref{fig:WCA_D}, where data following the critical power-law will fall on a straight line. While this holds for the single-energy case near the transition, where the critical asymptote becomes valid for roughly $\epsilon \lesssim 0.1$ as in the cherry-pit model, the canonical ensemble case shows a strong rounding.

Clearly, the canonical ensemble does not exhibit the critical dynamics of the Lorentz model, while the single-energy case does.
This can be explained by an averaging of the dynamics in the case of the canonical ensemble. In contrast to the cherry-pit model, the WCA system contains finite energy barriers. As a consequence, the available void space and its topology are a function of tracer energy, i.e. barriers which can be surmounted by fast tracers may not be passable for slower tracers. It will be shown in the following how this notion can be quantified with the help of a mapping of the system onto hard disks.

\paragraph{Hard-disk mapping}
The hard-disk mapping will yield an effective hard-disk interaction diameter $\sigma\eff$ for each tracer as a function of $n\WCA\reduced$ and its energy $E$ and will thus show that the dynamics in the WCA-disk system can be understood as an average over a distribution of effective Lorentz models.

What is needed is a mapping of the WCA-disk system onto an equivalent system with (overlapping) hard-disk obstacles and a point-like tracer.
In order for it to be useful, the mapping must conserve the topology of
the void space: open channels have to stay open and closed channels have to remain closed
under the mapping. Otherwise, the percolation transition of the void space would not be correctly mapped. While mappings such as the Barker--Henderson mapping, which was used to estimate the packing fraction of the matrix,  or a mapping using the Andersen-Weeks-Chandler approximation\cite{Schmiedeberg2011} can be very successful for the mapping of glassy systems, for example, they do not guarantee conservation of topology. Greater care is necessary, here.

In two dimensions, a channel in the void space is defined by two obstacles. 
The potential landscape in such a channel has the shape of a  saddle. 
A tracer is able to pass the channel if its energy matches or surpasses the
potential energy on the saddle point of the channel, i.e. at the
point exactly between the obstacles. 
In the presented mapping, the effective hard-core interaction distance $\sigma\eff$ between a given tracer and the obstacles is then
calculated as the closest distance between two obstacles forming a channel through which the tracer is barely able to pass.

For obstacles at a distance $2r$, the potential energy in the center of the channel is given by
$2V_{\alpha\beta}(r)$, \cref{eq:WCA} (the smoothing function $\Psi(r)$ can be neglected),
and a tracer of energy $E$ cannot pass the channel if $E < 2V_{\alpha\beta}(r)$.
Thus the topology of the accessible space is preserved if the soft obstacles are mapped to hard disks of radius $\sigma\eff$ (assuming a point tracer)
with the condition $E=2V_{\alpha\beta}(\sigma\eff)$, see \cref{fig:WCA_D_schematic}b. Explicitly,
\begin{equation}
 E = 8\eps\tracer
    \Biggl[
    \left(\frac{\sigma\tracer}{\sigma\eff}\right)^{12}  -
    \left(\frac{\sigma\tracer}{\sigma\eff}\right)^{6}
    + \frac{1}{4} \Biggr] ,
\end{equation}
which has two positive solutions for $\sigma\eff$. Only one of them respects the cutoff of the potential
$\sigma\eff \leq r\cut$,
\begin{equation}
 \sigma\eff = \left[\frac{1}{2} + \left(E/8\eps\tracer\right)^{1/2}\right]^{-1/6} \sigma\tracer.
 \label{eq:hardDiskDiameter}
\end{equation}
The reduced effective density of the matrix then reads
\begin{align}
 n\reduced\eff(E) := n\sigma\eff^2
 = n\WCA\reduced \left[\frac{1}{2} + \left(E/8\eps\tracer\right)^{1/2}\right]^{-1/3}.
 \label{eq:hardDiskDensity}
\end{align}

For the mapping to be successful, it has to correctly map the critical point as determined by the single-energy dynamics onto the percolation point of the matrix. From the dynamics, the critical point can be read off as
$n\WCA\reduced \approx 0.320$ where the simulation was performed at the tracer energy $E/\eps\matrx = 1.143$. Via the hard-disk mapping this corresponds to an effective hard-disk
critical radius of $(\sigma\eff)\crit/\sigma\matrx = 0.982$ and a critical hard-disc reduced
density $(n\reduced\eff)\crit = 0.268$. This fully agrees with the percolation threshold determined above via
Voronoi tesselation.

\paragraph{Energy-resolved dynamics}

The mapping clearly exposes that tracers with different energies experience matrices with different densities $n\reduced$. Thus, it is useful to consider the tracer dynamics as a function of tracer energy.
To this end, the total energy of each tracer was calculated at the beginning of the simulation and tracers with similar energies were grouped into bins of width $\Delta E/\eps\matrx = 0.1$. The MSD was then calculated for each tracer and averaged over each energy bin. Since the particles of each bin have approximately the same energy $E$ and same interaction range $\sigma\tracer$, their state can be uniquely expressed by the interaction diameter $\sigma\eff = \sigma\eff\bigl(\sigma\tracer, E\bigr)$.

The energy distribution of the tracers $p(E)$ corresponds to a distribution of effective densities $p(n\reduced\eff)$, which can be directly calculated via
\begin{align}
	p(n\reduced\eff) = - p\bigl(E(n\reduced\eff)\bigr) \,\frac{\diff E}{\diff n\reduced\eff}, \quad \text{for } E\geq 0,
\end{align}
and $p(n\reduced\eff) = 0$, else.
Note, that $E(n\reduced\eff)$ is given by the inversion of \cref{eq:hardDiskDensity} and that $\diff E / \diff n\reduced\eff$ is negative.

\begin{figure}
	\centering
		\includegraphics[width=\columnwidth]{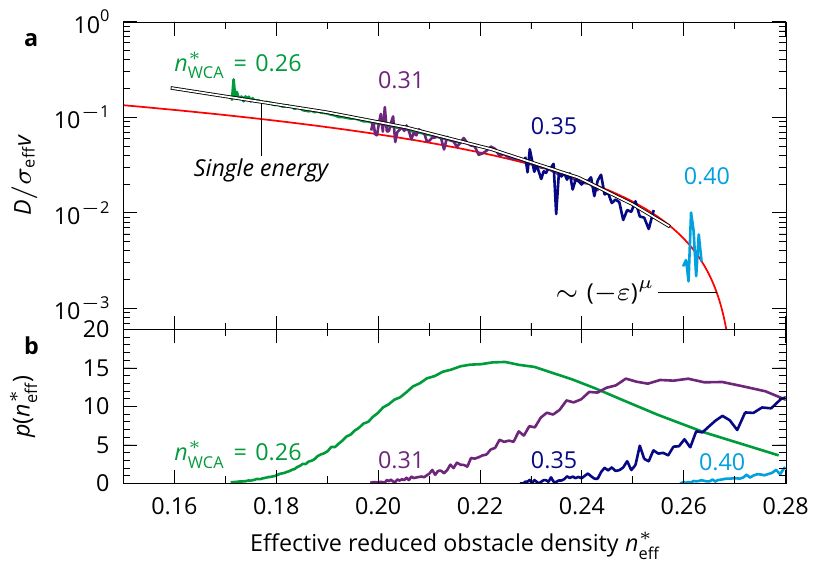}
    \caption{a) Master plot of diffusion coefficients $D$ in the WCA-disk
    system for a canonical ensemble of tracers resolved by their energy $E$.
    The white line shows $D$ for the single energy case of \cref{fig:WCA_D}. The red line indicates the
    critical asymptote from \cref{fig:WCA_D}.  b) Distribution of effective 
		reduced obstacle density in the same systems.}
	\label{fig:D_sigmadist}
\end{figure}

In \cref{fig:D_sigmadist}b, the distributions $p(n\reduced\eff)$ are shown for a range of $n\reduced\WCA$, with $p(E)$ directly taken from the simulation, see \cref{fig:energy_histogram_es}.
The diffusion coefficients calculated from the energy-resolved MSDs for the same systems are shown in \cref{fig:D_sigmadist}a.
To account for the trivial scaling with the microscopic time scale of the particles $t_o = \sigma\eff/\vel$ with the velocity $\vel$ of the particles, the diffusion coefficients have to be plotted rescaled as $D/(\sigma\eff \vel)$. The velocity $\vel$ for each energy was extracted from the short-time behavior of the MSD, $\msd(t;E) = \vel^2 t^2$. Without any further rescaling, the diffusion coefficient as a function of $n\reduced\eff$ falls onto a single master curve, which is in agreement with the single-energy data. As the percolation transition at $n\reduced\crit = 0.268$ is approached, the master curve approaches the critical behavior expected for the Lorentz model, $D\sim (-\epsilon)^\mu$. This demonstrates clearly that the hard-disk mapping is successful and that the long-time dynamics of single WCA-particles are compatible with the Lorentz model dynamics.

\begin{figure}
	\centering
		\includegraphics[width=\columnwidth]{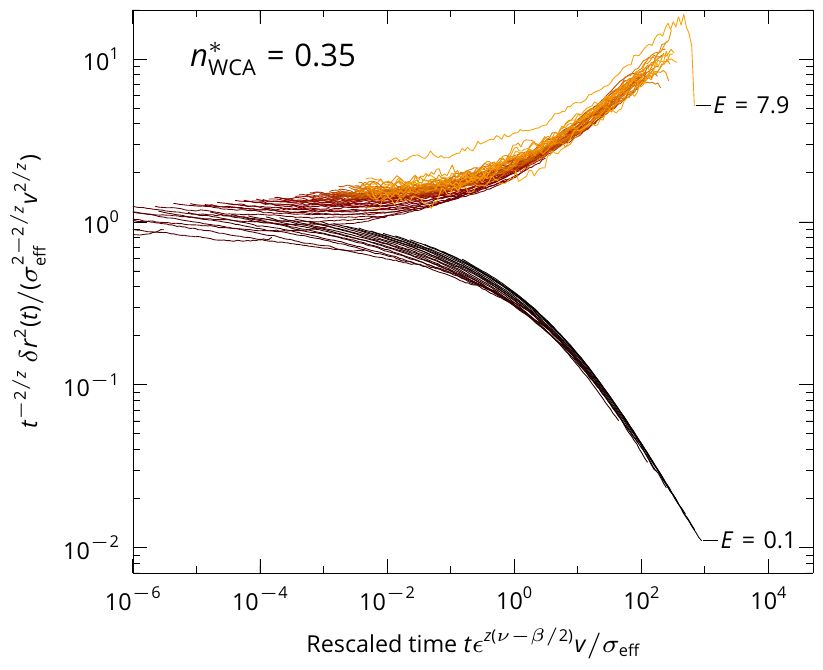}
	\caption{Master plot of energy-resolved mean-squared displacement in the WCA-disk system for a canonical ensemble of tracers at $n\WCA\reduced = 0.35$. The MSD is divided by
the critical asymptote $\sim t^{2/z}$ and shown as a function of rescaled time using the separation parameter $\epsilon := (n\reduced\eff - n\reduced\crit)/n\reduced\crit$ with $n\reduced\crit = 0.268$.
The energy $E$ of the MSDs increases from bottom to top and the MSDs with smallest and largest energy are annotated.}
	\label{fig:msd_energydist}
\end{figure}

Furthermore, it is possible to demonstrate that not only the diffusion coefficient but also the full dynamics satisfies the critical scaling of the Lorentz model when mapped onto hard-disks. For the Lorentz model, the MSD is expected to follow an asymptotic scaling which incorporates the regimes of regular and anomalous diffusion, as well as the localized regime into a single functional form,
\begin{align}
	\msd(t)  = t^{2/z} \mathcal \delta R_\pm^2(t/t_x) \,, \quad t_o \ll t,
\end{align}
with $t_x \sim \ell^{z} \sim |\epsilon|^{-z(\nu - \beta/2)}$ diverging at the transition~\cite{Hofling2013}. The scaling function $\delta R_-^2$ holds on the delocalized side of the transition, and $\delta R_+^2$ on the localized side. At long times, $t_x \ll t$, regular diffusion is recovered by $\delta R_-^2(t/t_x) \sim (t/t_x)^{1-2/z}$, while on the localized side $\delta R_-^2(t/t_x) \sim (t/t_x)^{2/z}$ holds. For $t_o \ll t \ll t_x$, both scaling functions tend to the same constant, representing the regime of anomalous diffusion.

Both scaling functions are displayed in \cref{fig:msd_energydist}, where the energy-resolved MSD at $n\reduced\WCA = 0.35$ is divided by the critical asymptote $t^{2/z}$ and time is rescaled appropriately. For this, the separation parameter $\epsilon$ was calculated from the effective interaction distance, $\epsilon = (n\reduced\eff - n\reduced\crit)/n\reduced\crit$ with $n\reduced\crit = 0.268$.
The collapse of the data onto the scaling functions is roughly as successful as in the overlapping Lorentz model \cite{Hofling2006} and can in principle be further improved by considering corrections to scaling \cite{Kammerer2008}.

\paragraph{Percolating fraction}
To quantify the rounding of the transition, it is instructive to calculate the fraction of tracers with an energy sufficiently high to allow for long-range transport. At a given $n\reduced\WCA$, this fraction corresponds to the percolation probability, $p\perc$, of the effective hard-disk system. Provided that $n\reduced\WCA$ is large enough that some tracers are on the localized side of the transition, $p\perc$ is
obtained as the integral over all subcritical states of $p(n\reduced\eff)$,
\begin{align}
	p\perc &:= \int_{0}^{n\reduced\crit} p(n\reduced\eff) \,\diff n\reduced\eff
  = \int_{E(n\reduced\crit)}^{E(0)} p(E) \,\diff E,
\end{align}
and $p\perc = 1$, otherwise.
At large densities $n\reduced\WCA$, only tracers with the largest energies are delocalized.
Then, it is reasonable to assume that $p(E) \approx A\beta \exp(-\beta E)$, e.g. from inspection of the inset of \cref{fig:energy_histogram_es}, with $A\gtrsim 1$ (If the approximation were meant to hold for all $E$, then due to normalization $A=1$ would hold exactly, but this would underestimate the probability distribution at large $E$). 
Furthermore, $E(n\reduced\eff\to 0) = +\infty$ holds. Thus, one finds
\begin{multline}
	p\perc  \approx A \exp \bigl(-\beta E(n\reduced\crit) \bigr) \\
     = A \exp \left\{ -8\beta \eps\tracer
    \left[
    \left(\frac{n\reduced\WCA}{n\reduced\crit}\right)^{6}  -
    \left(\frac{n\reduced\WCA}{n\reduced\crit}\right)^{3}
    + \frac{1}{4}
    \right] \right\}.
\end{multline}

The approximation of the energy distribution by an exponential
overestimates $p\perc$ at small densities, but the approximation should become exact
for large $n\reduced\WCA$. Therefore, $p\perc > 0$ holds for all finite $n\reduced\WCA$, but becomes exponentially suppressed at large densities.

\section{Summary and Conclusion}

We have performed simulations in two dimensions of particle transport in two models of porous media which represent systematic steps away from the standard overlapping Lorentz model towards realistic systems. In the Lorentz model, a percolation transition in the void space entails a localization transition in the dynamics with anomalous transport being a key signature.  Our systems allow testing which properties of porous media are necessary for anomalous transport and a localization transition.

In the cherry-pit model, the host matrix contains structural correlations which modify the structure of the void space. In the WCA model, interactions between tracer particles and the host particles are modeled with a purely repulsive and soft potential. This changes the topology of void space by introducing a potential energy landscape with finite barriers. The dynamics have been analyzed in terms of the mean-squared displacement and quantities derived from it.

In the cherry-pit model,
 we have determined the percolation threshold as a function of the obstacle packing fraction, which is a measure of structural correlations contained in the host matrix. For both a weakly and a strongly correlated system, the localization transition is observed coinciding with the percolation threshold and the dynamics is found to be compatible with the critical predictions for the Lorentz model. However, the convergence to the universal predictions is poor, with a possible origin being corrections to scaling. 
 
In the WCA model with a canonical ensemble of non-interacting tracer particles, a localization transition is observed, but the critical predictions do not apply, i.e.\ the transition is rounded. The behavior is similar to what has been observed in a quasi-twodimensional experiment recently~\cite{Skinner2013}.
The situation is clarified by a mapping of the WCA matrix onto hard-disks which reveals that the dynamics of each tracer can be fully mapped onto the Lorentz model as a function of its diameter and energy. This is confirmed by a scaling analysis of the dynamics as a function of tracer energy. The dynamics of the full system thus represent an energy-average over a distribution of effective Lorentz models.
As a consequence, in systems with soft potentials like WCA-disks, one can only observe the idealized Lorentz model scenario in a simulation where the energy of tracers can be precisely controlled and held constant over the whole simulation, i.e.\ \emph{only for Newtonian dynamics}. In Brownian dynamics or for interacting tracers, we expect the rounding of the transition to become more pronounced, as each tracer samples the full energy distribution over time.

\section{Acknowledgement} This work has been supported by the
Deutsche Forschungsgemeinschaft DFG via the  Research Unit FOR1394 ``Nonlinear Response to
Probe Vitrification''.

\bibliographystyle{rsc}
\bibliography{ConfinedIdealGas}

\end{document}